\documentclass[twocolumn,trackchanges]{aastex61}

\received{March 2, 2017}
\revised{March 28, 2017}
\accepted{March 30, 2017}

\submitjournal{AJ}

\shorttitle{The Algol Type Binary V Tri}
\shortauthors{Ren et al.}

\begin{document}

%%Title
\title{Photometric and Spectroscopic Observations of the Algol Type Binary V Triangle}

%%Authors
\correspondingauthor{J. N. Fu}
\email{jnfu@bnu.edu.cn}

\author[0000-0002-2037-2480]{A. B. Ren}
\affil{Department of Astronomy, Beijing Normal University, No. 19 Xinjiekouwai Street, Haidian District, Beijing 100875, China; renanbing@gmail.com}

\author{X. B. Zhang}
\affiliation{National Astronomical Observatories, Chinese Academy of Sciences, Beijing 100012, China; xzhang@bao.ac.cn}

\author{J. N. Fu}
\affil{Department of Astronomy, Beijing Normal University, No. 19 Xinjiekouwai Street, Haidian District, Beijing 100875, China; renanbing@gmail.com}

\author{Y. P. Zhang}
\affil{Department of Astronomy, Beijing Normal University, No. 19 Xinjiekouwai Street, Haidian District, Beijing 100875, China; renanbing@gmail.com}

\author{T. Q. Cang}
\affil{Department of Astronomy, Beijing Normal University, No. 19 Xinjiekouwai Street, Haidian District, Beijing 100875, China; renanbing@gmail.com}

\author{L. Fox-Machado}
\affiliation{Instituto de Astronom$\acute{\i}$a, Universidad Nacional Aut$\acute{o}$noma de M$\acute{e}$xico, Apartado Postal 877, Ensenada, Baja California, C.P. 22830 M$\acute{e}$xico;}

\author{C. Q. Li}
\affil{Department of Astronomy, Beijing Normal University, No. 19 Xinjiekouwai Street, Haidian District, Beijing 100875, China; renanbing@gmail.com}

\author{P. Khokhuntod}
\affil{Department of Astronomy, Beijing Normal University, No. 19 Xinjiekouwai Street, Haidian District, Beijing 100875, China; renanbing@gmail.com}

\author{Y. P. Luo}
\affiliation{Physics and Space Science College, China West Normal University, Nanchong 637002, China.}

%%Abstract
\begin{abstract} 
Time-series, multi-color photometry and high-resolution spectra of the short period eclipsing binary V Tri were obtained by observations. The completely covered light and radial velocity curves of the binary system are presented. All times of light minima derived from both photoelectric and CCD photometry were used to calculate the orbital period and new ephemerides of the eclipsing system. The analysis of $O-C$ diagram reveals that the orbital period is $0.58520481\ days$, decreasing at a rate of $dP/dt=-7.80\times10^{-8} d\ yr^{-1} $. The mass transfer between the two components and the light time-travel effect due to a third body could be used to explain the period decrease. However, a semidetached configuration with the less-mass component filling and the primary nearly filling each of their Roche lobes was derived from the synthesis of the light and radial velocity curves by using the 2015 version of the Wilson-Devinney code. We consider the period decrease to be the nonconservative mass transfer from the secondary component to the primary and the mass loss of the system, which was thought to be an EB type while it should be an EA type (semi-detached Algol-type) from our study. The masses, radii and luminosities of the primary and secondary are $1.60\pm0.07 M_\odot$, $1.64\pm0.02 R_\odot$, $14.14\pm0.73 L_\odot$ and $0.74\pm0.02 M_\odot$, $1.23\pm0.02 R_\odot$, $1.65\pm0.05 L_\odot$, respectively.
\end{abstract}


%%--Section1
\section{Introduction} 
\label{sect:1}

V Tri (GSC 02293-01403, SV* HV 3347), short-period eclipsing binary system, was discovered by Miss Leavitt \citep{leav13} from Harvard plates according to the discordance of the resulting magnitude. The type of this system is classified as the Algol type with a magnitude range from $10^{m}.5$ and $11^{m}.8$. \citet{hoff19} identified its $\beta$ Lyrae type nature through photographical photometry (magnitude: $10^{m}.6$ $\sim$  $11^{m}.5$). Photographic light curve of the system were derived by \citet{jord29} firstly, the range of magnitude that published is greater than the results of \citet{hoff19}, the secondary minimum is less deep, however. \citet{cann34} determined the spectral type of V Tri is A3. The geometric and physical parameters for this binary system were determined using the iterative method by \citet{bran80}, they revised the spectral types as A3+F6 for both components and also classified it as $\beta$ Lyrae type. \citet{giur83} marked V Tri as an ``a" systems which means a non-contact configuration, but some authors were inclined to a near-contact configuration (\cite{budd84}; \cite{shaw94}; \cite{gray94}). In recent study, \citet{guro06} derived the physical parameters of the system from $BVR$ light curves using the Mode 2 (detached configuration) and Mode 5 (semi-detached configuration with the the secondary component accurately filling its limiting lobe) of Wilson-Devinney (WD) 2003 code.

A mount of light minimum times of V Tri have been published by many amateurs and professional astronomers with visual, photographic, photoelectric, and CCD observations in the database of times of minima and maxima\footnote{http://var2.astro.cz/ocgate/}. The ephemeris of the system have been updated gradually using the light minima (\cite{hoff19}; \cite{jord29}; \cite{wood63}). The peculiar period behavior of this systems were discussed by \cite{gray94} based on light minimum timings from 1902 to 1992. Furthermore, \citet{guro06} analyzed the period of V Tri and attributed the $O-C$ (observed minus calculated period) variations of all minima to the light-time effect caused by a third body.

In order to study the period variation and determine a set of precise absolute parameters of V Tri, we carried out both photometric and spectroscopic observations for this binary system. Precision $BV$ band light curves in a phase, 6 new light minima and 10 radial velocities (RVs) of the primary component have been obtained at the different observing stations in recent years. In this paper, we shall first introduce the observations and image-processing progress in Section \ref{sect:2}. New ephemerides and period variation analysis based on the photoelectric and CCD minimum times are given in Section \ref{sect:3}. In Section \ref{sect:4}, we derived the photometric solution of the system based on the high accurate light curves and RVs of the primary star using the 2015 version WD code. The summary and discussion about this binary system will be presented in the last section.

%%--Section 2
\section{Observations and Data Processing}  
\label{sect:2}

The photometric observations for V Tri were carried out on two seasons to obtained more times of light minima and complete light curves in a phase using the 85 cm reflector telescope at the Xinglong Station of the NAOC\footnote{National Astronomical Observatories, Chinese Academy of Sciences.} in December 2003 and October 2007, respectively. In the first season, the data were collected with an AP7P $512\times512$ CCD camera which provides a field of view of about $6' \times6'$ with an image scale of about $0''.7$ per pixel. A single Johnson $V$ filter was used and the exposure time was 60 second for each exposure. A total number of about 400 useful frames were obtained on two nights. In the second season, all CCD frames were collected using a PI MicroMAX 1024 BFT CCD camera with a field of view of about $16'.5\times16'.5$. The pixel scale is $0''.96$ per pixel. Two standard Johnson-Cousin-Bessell filters in $B$ and $V$ bands were used to record the variations of light curves on the three days, the typical exposure times were set at 30s or 15s for $B$ band and 15s or 8s for $V$ band measurements depending on the weather condition. The total of useful frames are 2125 in $B$ and 2122 in $V$ band. Since these data record a entire light curve of the eclipsing system for each filter,   the early photometric data were used only to determine the times of minima. Finally, we obtained six light minima and complete $B$ and $V$ band light curves of V Tri. Detail journal of observation is given in Table \ref{tab1}.

\begin{figure}  
\centering
\includegraphics[width=0.48\textwidth, angle=0]{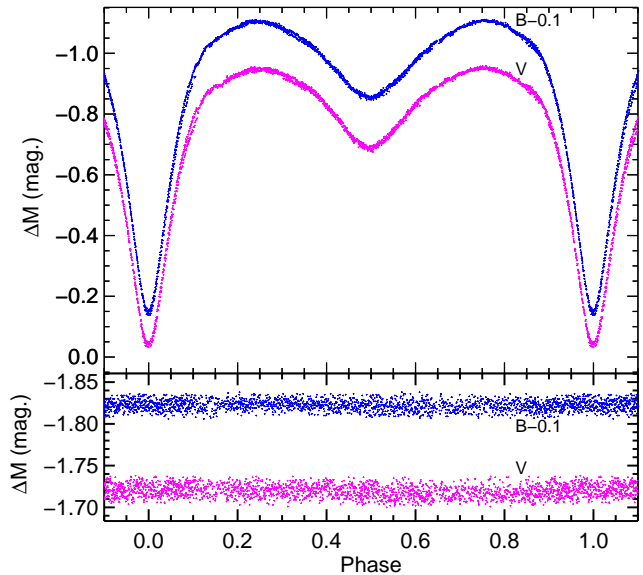}
\caption{The differential light curves of V Tri in the $B$ and $V$ band. The differential magnitude between the Comparison star and Check star are plotted in the down panel.}
\label{fig1}
\end{figure}

All CCD images were reduced preliminarily with the standard process of the IRAF\footnote{Image Reduction and Analysis Facility}/CCDRED package including subtract the bias and dark, and divide flat-fields from the object frames, and then the instrument magnitudes of the stars were extracted from these images using the aperture photometry of IRAF/DAOPHOT package \citep{stet87}. All relatively bright stars with good seeing in the viewing field were selected out as reference candidate stars to perform differential photometry. Finally, GPM 22.913124+30.390996 and 2MASS J01321631+3022599 were employed as the comparison and check stars, respectively. The main information of the object, comparison and check star, taken from the web of Aladin sky atlas, are given in Table \ref{tab2}. The observing time has been transformed into Heliocentric Julian Day (HJD) to obtained the phased light curve of the object as given in the up panel of Figure \ref{fig1}. The magnitude difference between the comparison and the check was performed to be stable within 0.03 magnitude during the observations in $B$ and $V$ band as shown in the bottom panel.

\begin{figure} 
\centering
\includegraphics[width=0.48\textwidth, angle=0]{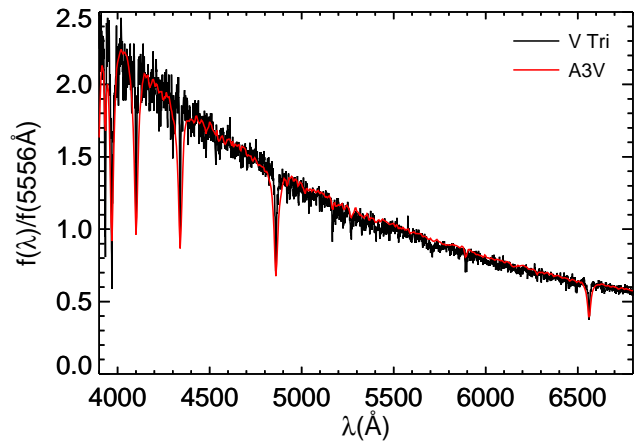}
\caption{The black curve represent the spectrum of V Tri, the spectrum of A3V from the stellar spectral flux library is indicated by the red curve.}
\label{fig2}
\end{figure}

The spectroscopic observations for V Tri were carried out with the 2.12 m telescope at the Observatorio Astron\'emico Nacional on the Sierra San Pedro M\'artir (OAN-SPM) in M\'exico on November 03 and 05, 2015. We used a $2048 \times 2048$ E2V CCD-4240 to collected the high resolution (the maxima resolution is $R=18,000$ at 5,000 \AA) echelle spectra at the slit size $1''$. The spectral range coverage from 3,800 to 7,100 \AA. We finished the preliminary spectrum process including bias and flat calibration in the IRAF/CCDRED package. The cosmic rays were effectively removed from the object frames using the STSDAS/lacos\_spec package in IRAF. Then the data reduction was mainly performed using the IRAF/ECHELLE package to obtain the normalized continuous spectrum as shown in Figure \ref{fig2}. The spectral line of the binary system are agree well with the spectral energy distribution for A3V from the stellar spectral library by \cite{pick98}.  Due to the large brightness difference between the two components of the binary system, we only calculated the RVs of the primary star from the processed spectra according to the formula 
%%equation 1
\begin{equation} 
RV = C \cdot \frac{\Delta \lambda}{\lambda}
\label{eq1}
\end{equation} 
where $C$ is the speed of light, $\Delta \lambda$ is the wavelength difference between the object and radial-velocity standard star at the center position of the same absorption line. $\lambda$ is the standard center wavelength of the absorption line. Finally, we convert the $RV$ to the heliocentric radial velocity by using the task $bcvcorr$ in the IRAF/RVSAO package. Each mean radial velocity value with the statistical error is calculated from the observed spectral data as listed in Table \ref{tab3}.
%%Table1 
\begin{table*}[!t] 

%Table 1
\begin{center}
\caption{Journal of the Time-series CCD Photometric and Spectroscopic Observations for V Tri.}\label{tab1}
\def\temptablewidth{0.67\textwidth}
{\rule{\temptablewidth}{0.5pt}}
\begin{tabular*}{\temptablewidth}{@{\extracolsep{\fill}}  lcrrrc}
\hline
Date (UT)        & Observatory   & Telescop   & Filter   & $N_{obs}$ & Observer \\
\hline                                               
\multicolumn{2}{ l }{Photometric Observations}  &       &                   &            \\
\cline{1-2}
2003 Dec. 25  & BAO               & 85cm         & V        &  297           & XBZ     \\  
2003 Dec. 26  & BAO               & 85cm         & V        &  297           & XBZ     \\  
2007 Oct. 12   & BAO               & 85cm         & BV     &  914           & XBZ     \\ 
2007 Oct. 13   & BAO               & 85cm         & BV     & 1218          & XBZ     \\
2007 Oct. 14   & BAO               & 85cm         & BV     & 2115          & XBZ     \\
\hline
\multicolumn{2}{ l }{Spectroscopic Observations}  & & & \\
\cline{1-2}
2015 Nov. 03  & SPM               & 2.12m        &           &    3            & LFM,TQC \\
2015 Nov. 05  & SPM               & 2.12m        &           &    7            & LFM,TQC \\
\end{tabular*}
{\rule{\temptablewidth}{0.5pt}}                                                                                                  

\begin{minipage}[t]{0.67\textwidth}                                                                                              
\textbf{Notes.} XBZ: Xiao-Bin Zhang; LFM: Lester Fox-Machado; TQC: Tian-Qi Cang.
\end{minipage}                                                                                                                   

\end{center}  
\end{table*}

%%Figure1

%%Figure2

%%Table2
\begin{table*}[!t] 

%Table 2
\begin{center}
\caption{Main Feature Parameters of the Variable, Comparison and Check Stars.}\label{tab2}
\def\temptablewidth{0.77\textwidth}
{\rule{\temptablewidth}{0.5pt}}
\begin{tabular*}{\temptablewidth}{@{\extracolsep{\fill}}  lllllll}
\hline
Star                & RA                                        & Dec                                  & $B$     & $V$       & $B-V$    & Sp\_type \\
                       & (ep=J2000)                          & (ep=J2000)                      & (mag)  & (mag)    & (mag)     &                \\
\hline
V Tri               & $01^{h}31^{m}47^{s}.103$   & $+30^{\circ} 22' 01''.57$  & 11.32   & 10.96    & 0.36        & A3           \\    
Comparison   & $01^{h}31^{m}39^{s}.158$   & $+30^{\circ} 23' 27''.76$  & 11.3     &              &                &                \\    
Check            & $01^{h}32^{m}16^{s}.32 $    & $+30^{\circ} 22' 59''.9  $  & 13.7     & 13.4      & 0.3          &                \\    
\end{tabular*}
{\rule{\temptablewidth}{0.5pt}}
\end{center}
\end{table*}

%%Table3
\begin{table*}[!t] 

%Table 3
\begin{center}
\caption{Heliocentric Radial Velocities of the Primary Component for V Tri.}\label{tab3}
\def\temptablewidth{0.67\textwidth}
{\rule{\temptablewidth}{0.5pt}}
\begin{tabular*}{\temptablewidth}{@{\extracolsep{\fill}}  lcr}
\hline
HJD                 & Phase     & RV                \\
(2,400,000+)   &                &  km/s             \\
\hline
57329.8787    &  0.6448   &   18$\pm$5    \\
57329.9087    &  0.6962   &   36$\pm$5    \\
57329.9535    &  0.7727   &   45$\pm$7    \\
57331.7023    &  0.7610   &   45$\pm$7    \\
57331.7255    &  0.8008   &   44$\pm$8    \\
57331.8449    &  0.0048   &  -74$\pm$10  \\
57331.8754    &  0.0568   & -120$\pm$10 \\
57331.9108    &  0.1173   & -134$\pm$9   \\
57331.9333    &  0.1559   & -144$\pm$9   \\
57331.9667    &  0.2128   & -158$\pm$7   \\
\end{tabular*}
{\rule{\temptablewidth}{0.5pt}}
\end{center}
\end{table*}

%% --Section3
\section{New Ephemerides and Period Variation}   
\label{sect:3}

As an interesting study target V Tri, many amateur and professional astronomers observed the light minimum times using the visual, photographic, photoelectric or CCD methods. In our observations, 6 light minima of this binary system were recorded on 5 nights. The times of the light minima were determined by fitting the light curves with a quadratic function, their errors were estimated using Monte Carlo method. Moreover, 75 minimum times that measured by photoelectric (pe) and CCD photometry were picked out from all available light minima to study the orbital period variation of the system. All of 81 minima and their errors (if available) are listed in the first two columns of Table \ref{tab4}. 
Based on the latest linear ephemeris
\begin{eqnarray} %%equation 2
  Min.I(HJD)&=&2,452,490.44221\pm0.00052 \nonumber \\
&+&(0.585205686\pm0.000000024)E
\label{eq2}
\end{eqnarray}
was published by \cite{guro06} from photoelectric and CCD measurements. Using the classical $O-C$ method, we calculated the revised linear and quadratic ephemerides for V Tri as following.
\begin{eqnarray} %%equation 3
  Min.I(HJD)&=&2,454,387.0918\pm0.0001 \nonumber \\
                   &+&(0.58520500\pm0.00000004)E
\label{eq3}
\end{eqnarray}
\begin{eqnarray} %%equation 4
  Min.I(HJD)&=&2,454,387.0924\pm0.0001 \nonumber \\
          &+&(0.58520481\pm0.00000004)E \nonumber \\
          &-&(6.24\pm0.69)\times10^{-11}E^2
\label{eq4}
\end{eqnarray}

\begin{figure}  %%Figure3
\centering
\includegraphics[width=0.48\textwidth, angle=0]{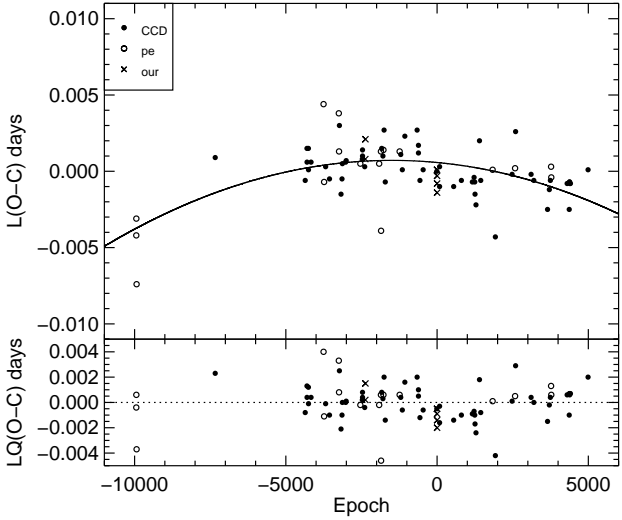}
\caption{Orbital period analysis for V Tri.}
\label{fig3}
\end{figure}

The $O-C$ linear and quadratic residuals for the times of light minimum were calculated as listed in the sixth and seventh column of Table \ref{tab4}, respectively. The $O-C$ diagram of period analysis is displayed in Figure \ref{fig3}, the solid line represents the quadratic  approximation refer to the quadratic ephemeris. Although there are no more reliable light minimum times were published from HJD2448573.6579 to HJD2451835.5974 (about between 1992 and 1999), the period analysis produces a good fitting for these photoelectric and CCD minimum times and yields a new period $P=0.58520481\ days$ that slightly shorter than those obtained by Wood \& Forbes (0.58520771 days; \citeyear{wood63}), Gray et al. (0.5852060 days; \citeyear{gray94}), Kreiner et al. (0.58520570 days; \citeyear{krei01}) and G\"{u}rol et al. (0.58520569 days; \citeyear{guro06}) but it's obviously longer than that of Hoffmeister (0.5851993 days; \citeyear{hoff19}). The downward parabola, as shown in the Figure \ref{fig3}, suggested that the orbital period of the system was undergoing a continuous period decrease in the past few decades. The rate of continuous period decrease was calculated about $d\emph{P}/d\emph{E}=-1.25\times10^{-10}d\ cycle^{-1}$ or $d\emph{P}/d\emph{t}=-7.80\times10^{-8}d\ yr^{-1}$ from the quadratic ephemeris. The period decrease maybe indicated that the mass transfer from the more-massive component to the less-massive one. 

\startlongtable %%Table4
\begin{deluxetable*}{lllcrrrr}

%Table 4
\tablecaption{The Photoelectric and CCD Times of Light Minima of V Tri along with the Residuals Computed from the Quadratic Ephemerides.\label{tab4}}
\tablewidth{700pt}
\tabletypesize{\scriptsize}
\tablehead{
\colhead{HJD}          & \colhead{Error}  & \colhead{Method} & \colhead{Min} &  \colhead{E} & \colhead{L(O-C)}   &  \colhead{LQ(O-C)}  & \colhead{Ref.} \\
\colhead{(2,400,000+)} &  \colhead{}    & \colhead{}             &  \colhead{}      &  \colhead{}    & \colhead{days}      &  \colhead{days}        &  \colhead{}
} 
\startdata
 48567.8091  &              & pe        & I   &-9944.0 & -0.0042  &  -0.0004  & 1  \\   
 48568.6880  &              & pe        & II  &-9942.5 & -0.0031  &   0.0006  & 1  \\   
 48573.6579  &              & pe        & I   &-9934.0 & -0.0074  &  -0.0037  & 1  \\   
 50094.6140  &              & CCD    & I   &-7335.0 &  0.0009  &   0.0023  & 2  \\   
 51835.5974  & 0.0004  & CCD    & I   &-4360.0 & -0.0006  &  -0.0008  & 3  \\   
 51869.5414  & 0.0001  & CCD    & I   &-4302.0 &  0.0015  &   0.0013  & 3  \\   
 51870.7109  & 0.0009  & CCD    & I   &-4300.0 &  0.0006  &   0.0004  & 3  \\   
 51897.6312  & 0.0002  & CCD    & I   &-4254.0 &  0.0015  &   0.0012  & 3  \\   
 51899.3855  & 0.0020  & CCD    & I   &-4251.0 &  0.0001  &  -0.0001  & 4  \\   
 51948.5432  & 0.0002  & CCD    & I   &-4167.0 &  0.0006  &   0.0004  & 3  \\   
 52190.5292  & 0.0026  & pe        & II  &-3753.5 &  0.0044  &   0.0040  & 5  \\   
 52201.3504  & 0.0005  & pe        & I   &-3735.0 & -0.0007  &  -0.0011  & 5  \\   
 52230.6117  & 0.0001  & CCD    & I   &-3685.0 &  0.0003  &  -0.0001  & 3  \\   
 52305.5171  & 0.0002  & CCD    & I   &-3557.0 & -0.0005  &  -0.0010  & 3  \\   
 52485.4720  & 0.0024  & pe        & II  &-3249.5 &  0.0038  &   0.0033  & 5  \\   
 52490.4437  & 0.0003  & pe        & I   &-3241.0 &  0.0013  &   0.0008  & 5  \\   
 52497.4679  & 0.0003  & CCD    & I   &-3229.0 &  0.0030  &   0.0025  & 6  \\   
 52528.7718  & 0.0002  & CCD    & II  &-3175.5 & -0.0015  &  -0.0021  & 7  \\   
 52547.7920  & 0.0010  & CCD    & I   &-3143.0 & -0.0005  &  -0.0010  & 7  \\   
 52554.8155  & 0.0001  & CCD    & I   &-3131.0 &  0.0005  &  -0.0000  & 7  \\   
 52618.6029  & 0.0002  & CCD    & I   &-3022.0 &  0.0006  &   0.0000  & 3  \\   
 52628.5515  & 0.0001  & CCD    & I   &-3005.0 &  0.0007  &   0.0001  & 3  \\   
 52902.4272  & 0.0001  & pe        & I   &-2537.0 &  0.0005  &  -0.0002  & 8  \\   
 52940.4659  & 0.0010  & CCD    & I   &-2472.0 &  0.0008  &   0.0002  & 4   \\   
 52944.2703  & 0.0011  & CCD    & II  &-2465.5 &  0.0014  &   0.0008  & 9   \\   
 52944.5625  & 0.0005  & CCD    & I   &-2465.0 &  0.0010  &   0.0004  & 9   \\   
 52986.6965  &              & CCD    & I   &-2393.0 &  0.0003  &  -0.0004  & 10 \\  
 52998.9864  & 0.0001  & CCD    & I   &-2372.0 &  0.0008  &   0.0002  & 11 \\  
 53001.0359  & 0.0003  & CCD    & II  &-2368.5 &  0.0021  &   0.0015  & 11 \\  
 53266.4247  & 0.0001  & pe        & I   &-1915.0 &  0.0005  &  -0.0002  & 12 \\ 
 53300.3674  & 0.0001  & pe        & I   &-1857.0 &  0.0013  &   0.0006  & 12 \\ 
 53303.5809  & 0.0003  & pe        & II  &-1851.5 & -0.0039  &  -0.0046  & 12 \\ 
 53315.5830  &              & CCD    & I   &-1831.0 &  0.0015  &   0.0008  & 10 \\  
 53339.5759  &              & CCD    & I   &-1790.0 &  0.0010  &   0.0003  & 10 \\  
 53349.2321  & 0.0007  & pe        & II  &-1773.5 &  0.0014  &   0.0006  & 12 \\  
 53361.2302  & 0.0015  & CCD    & I   &-1753.0 &  0.0027  &   0.0020  & 4   \\   
 53384.6350  &              & CCD    & I   &-1713.0 & -0.0007  &  -0.0014  & 10 \\  
 53662.6093  & 0.0012  & pe        & I   &-1238.0 &  0.0013  &   0.0006  & 13 \\  
 53684.2617  & 0.0002  & CCD    & I   &-1201.0 &  0.0011  &   0.0004  & 14 \\  
 53714.6914  &              & CCD    & I   &-1149.0 &  0.0001  &  -0.0006  & 10 \\  
 53763.2656  & 0.0008  & CCD    & I   &-1066.0 &  0.0023  &   0.0016  & 4   \\   
 54000.5666  & 0.0002  & CCD    & II  &  -660.5 &  0.0027  &   0.0020  & 4   \\   
 54026.3146  & 0.0027  & CCD    & II  &  -616.5 &  0.0017  &   0.0010  & 15 \\  
 54026.6067  & 0.0002  & CCD    & I   &  -616.0 &  0.0012  &   0.0005  & 15 \\  
 54055.2800  & 0.0002  & CCD    & I   &  -567.0 & -0.0006  &  -0.0012  & 16 \\  
 54117.3124  & 0.0001  & CCD    & I   &  -461.0 &  0.0001  &  -0.0006  & 17 \\  
 54381.5324  & 0.0009  & CCD    & II  &      -9.5 &  0.0000  &  -0.0005  & 18 \\   
 54386.2141  & 0.0003  & CCD    & II  &      -1.5 &  0.0001  &  -0.0005  & 11 \\   
 54387.0915  & 0.0001  & CCD    & I   &       0.0 & -0.0003  &  -0.0009  & 11 \\   
 54387.9682  & 0.0005  & CCD    & II  &       1.5 & -0.0014  &  -0.0020  & 11 \\   
 54388.2614  & 0.0001  & CCD    & I   &       2.0 & -0.0008  &  -0.0014  & 11 \\   
 54435.6628  & 0.0001  & CCD    & I   &     83.0 & -0.0010  &  -0.0016  & 19 \\  
 54435.6641  & 0.0001  & CCD    & I   &     83.0 &  0.0003  &  -0.0003  & 19 \\  
 54707.7832  & 0.0001  & CCD    & I   &   548.0 & -0.0010  &  -0.0014  & 20 \\ 
 54857.3034  & 0.0011  & CCD    & II  &   803.5 & -0.0006  &  -0.0010  & 21 \\
 55069.4402  & 0.0002  & CCD    & I   & 1166.0 & -0.0007  &  -0.0009  & 22 \\   
 55102.5045  & 0.0016  & CCD    & II  & 1222.5 & -0.0004  &  -0.0007  & 22 \\   
 55115.6713  & 0.0001  & CCD    & I   & 1245.0 & -0.0007  &  -0.0010  & 23 \\   
 55120.9374  & 0.0002  & CCD    & I   & 1254.0 & -0.0015  &  -0.0017  & 23 \\   
 55139.6633  & 0.0003  & CCD    & I   & 1286.0 & -0.0022  &  -0.0024  & 23 \\   
 55207.2586  & 0.0006  & CCD    & II  & 1401.5 &  0.0020  &   0.0018  & 24 \\   
 55231.5420  & 0.0001  & CCD    & I   & 1443.0 & -0.0006  &  -0.0008  & 23 \\   
 55461.5283  & 0.0028  & pe        & I   & 1836.0 &  0.0001  &   0.0001  & 25 \\   
 55514.4850  & 0.0016  & CCD    & II  & 1926.5 & -0.0043  &  -0.0042  & 22 \\   
 55837.8148  & 0.0001  & CCD    & I   & 2479.0 & -0.0002  &   0.0001  & 26 \\   
 55896.3357  & 0.0010  & pe        & I   & 2579.0 &  0.0002  &   0.0005  & 27 \\   
 55905.7014  & 0.0004  & CCD    & I   & 2595.0 &  0.0026  &   0.0029  & 28 \\   
 56205.9088  & 0.0004  & CCD    & I   & 3108.0 & -0.0002  &   0.0004  & 29 \\   
 56258.5768  & 0.0001  & CCD    & I   & 3198.0 & -0.0006  &   0.0000  & 30 \\   
 56523.9654  & 0.0003  & CCD    & II  & 3651.5 & -0.0025  &  -0.0015  & 31 \\   
 56558.7864  & 0.0003  & CCD    & I   & 3711.0 & -0.0012  &  -0.0002  & 31 \\   
 56578.6839  & 0.0001  & CCD    & I   & 3745.0 & -0.0006  &   0.0004  & 29 \\   
 56592.4372  & 0.0025  & pe        & II  & 3768.5 &  0.0003  &   0.0013  & 32 \\   
 56596.5329  & 0.0048  & pe        & II  & 3775.5 & -0.0004  &   0.0006  & 32 \\   
 56902.8873  & 0.0001  & CCD    & I   & 4299.0 & -0.0008  &   0.0006  & 29 \\   
 56943.5574  & 0.0052  & CCD    & II  & 4368.5 & -0.0025  &  -0.0010  & 33 \\   
 56949.4112  & 0.0069  & CCD    & II  & 4378.5 & -0.0007  &   0.0007  & 33 \\   
 56949.7038  & 0.0001  & CCD    & I   & 4379.0 & -0.0007  &   0.0007  & 29 \\   
 56952.6297  & 0.0001  & CCD    & I   & 4384.0 & -0.0008  &   0.0006  & 29 \\   
 56966.6747  & 0.0001  & CCD    & I   & 4408.0 & -0.0008  &   0.0007  & 29 \\   
 57307.8501  &              & CCD    & I   & 4991.0 &  0.0001  &   0.0020  & 34 \\   
\enddata
\tablecomments{Ref. \citealp[1,][]{gray94}; \citealp[2,][]{bald97}; \citealp[3,][]{bald03}; \citealp[4,][]{brat07}; \citealp[5,][]{muye03}; \citealp[6,][]{demi03}; \citealp[7,][]{nels03}; \citealp[8,][]{hubs05a}; \citealp[9,][]{maci04}; \citealp[10,][]{bald06}; 11, Present Study; \citealp[12,][]{hubs05b}; \citealp[13,][]{hubs06}; \citealp[14,][]{zejd06}; \citealp[15,][]{hubs07}; \citealp[16,][]{dogr07}; \citealp[17,][]{dogr09}; \citealp[18,][]{hubs08}; \citealp[19,][]{samo08a}; \citealp[20,][]{samo08b}; \citealp[21,][]{hubs10}; \citealp[22,][]{brat11}; \citealp[23,][]{samo10}; \citealp[24,][]{goka12}; \citealp[25,][]{hubs11}; \citealp[26,][]{samo12}; \citealp[27,][]{hubs12}; \citealp[28,][]{diet12}; \citealp[29,][]{samo15b}; \citealp[30,][]{samo13b}; \citealp[31,][]{samo13a}; \citealp[32,][]{hubs14}; \citealp[33,][]{hubs15}; \citealp[34,][]{samo15a}.}
 
\end{deluxetable*}

\begin{figure}  %%Figure4
\centering
\includegraphics[width=0.48\textwidth, angle=0]{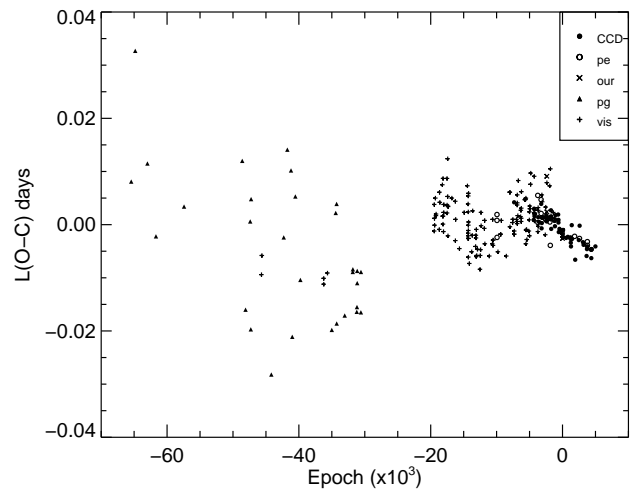}
\caption{The $O-C$ variation of V Tri in reference to Eq. \ref{eq2}.}
\label{fig4}
\end{figure}

A third body may be as the cause of period change in Algol-type systems \citep{hoff06}. The period variation of the V Tri have been attributed to the light-time effect by \cite{guro06}, they assumed a third body cause the sinusoidal variation in the orbital period of the eclipsing binary star system. Including our 6 CCD light minima, we collected all available light minima from literatures to find out how the change of orbital period will be in the last few decades. The $O-C$ variation of all light minimum times is shown in Figure \ref{fig4}. We can find a sinusoidal-like $O-C$ pattern after Epoch $> -20\times10^3$, which may be indicated that this system consists of the binary V Tri orbited by a distant third body. One visual value (2429217.4780) are excluded from the figure due to its great deviation compared with other data.

%% --Section4
\section{Light curves and Photometric Solution}
\label{sect:4}
All the observations have been transferred into phases by using the newly derived quadratic ephemeris as shown in Figure \ref{fig1}. Unlike that obtained by \cite{jord29}, the obviously symmetry of light curve is shown at the maximum but no so-called O'Connell effect on light curves as pointed out by \cite{guro06}. Agree with the primal classified type \citep{leav13}, the general feature of the light curves is belongs to that of Algol-type binary system with equal light maximum and minimum. The magnitude variation is derived to be about 0.96 and 0.89 mag for the primary eclipse and about 0.24 and 0.27 mag for the secondary one in $B$ and $V$ band, respectively. 

The photometric solution of the binary system was carried out to derive the system parameters with the complete light curves in $B$ and $V$ band and RVs of the primary component. The 2015 version of the Wilson-Devinney (W-D) code \citep{wils90, wils79, wils71, wils16} with the Kurucz atmospheres \citep{kuru93} was applied to the numerical light curve analysis. The nonlinear limb-darkening law with the logarithmic form was added in the analysis progress. We defined the massive component as Star 1 and the less massive one as Star 2 in the following photometric analysis. In order to derive the photometric solution, we initially assumed some input parameters and kept them constant during running W-D code. According to the spectral type determined from the spectrum (Figure \ref{fig2}), and through the calibration of \cite{cox00}, the temperature of star 1 was modified as $T_{1}=8750$ K. The gravity-darkening exponents were given as $g_{1}=1.0$ for Star 1, $g_{2}=0.32$ for Star 2 and the bolometric albedos were taken as $A_{1}=1.0$, $A_{2}=0.5$ \citep{wils16}. The initial bolometric ($X_{1}$, $X_{2}$, $Y_{1}$, $Y_{2}$) and monochromatic ($x_{1}$, $x_{2}$, $y_{1}$, $y_{1}$) limb-darkening coefficients of the components were adopted from \cite{vanh93}. The adjustable parameters are listed as follows: the phase shift, ${\o}_{0}$, the orbital inclination of the binary system, $i$, the mean temperature of Star 2, $T_{2}$, the surface potentials $\Omega_{1}$ and $\Omega_{2}$ for both components, the monochromatic luminosities of Star 1, $L_{1B}$ and $L_{1V}$, and the mass ratio, $q=m_{2}/m_{1}$.

\begin{figure}  %%Figure5
\centering
\includegraphics[width=0.48\textwidth, angle=0]{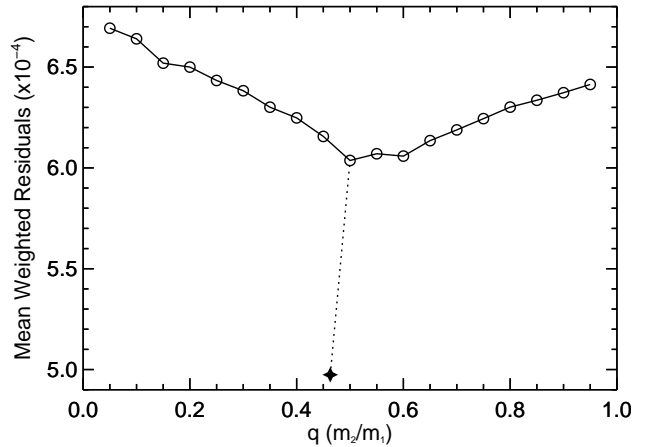}
\caption{The relation that the mean weighted residuals versus the assumed $q$. The best-fit photometric solution converged at $q=0.463$ as labeled with star symbol.}
\label{fig5}
\end{figure}

\cite{bran80} and \cite{bond96} published a different mass ratio 0.60 and 0.34 in their works, respectively. The latest photometrical study for V Tri, \cite{guro06} searched two mass ratio for detached configuration (0.508) and semi-detached configuration (0.458) with the q-searched method. 
We assumed a probable mass ratio range from 0.05 to 0.95, then a series of test values was used to search for a reliable mass ratio of Star 2 and 1. We run the differential corrections main program (DC) from mode 2 and the program was then turned into the state of mode 5. A set of convergent numerical solutions corresponding to each assumed mass ratio $q$ was computed out after many iterations of the adjustable parameters. The relation between the mean weighted residuals and the assumed $q$ is described with open circles in Figure \ref{fig5}. A photometric mass ratio was considered at around $q=0.5$ from the result of the $q$-search. Then we set $q$ as an adjustable parameter and ran the DC program again to compute out a more reliable value. After enough iterations, the best-fit photometric solution converged at $q=0.463\pm0.001$ (the star symbol in Figure \ref{fig5}) for semi-detached configuration. Next, the orbital semi-major axis $a$ was adjusted. Finally, we derived a set of parameters of photometric analysis for the eclipsing binary system as listed in Table \ref{tab5}.

\begin{deluxetable}{lll}  %%Table5

%Table 5
\tablecaption{Photometric Solution for V Tri.\label{tab5}}
\tablewidth{700pt}
\tabletypesize{\scriptsize}
\tablehead{ \colhead{Parameter}   &   \colhead{Best-fit Value}   &  \colhead{Formal Error}}
\startdata
$a(R_\odot)$                              &  3.91                 &  0.10             \\
$q=m_{2}/m_{1}$                       & 0.463                &  0.001           \\
$T_{1}(K)$                                 & 8750                 & $b$                \\
$T_{2}(K)$                                 & 5902                 & $\pm6$          \\
$g_{1}, g_{2}$                            & 1.0, 0.32           & $b$                \\
$i$                                              & 84.19                & $\pm0.05$     \\
$A_{1}, A_{2}$                           & 1.0, 0.5              & $b$                \\
$\Omega_{1}$                           & 2.9375               & $\pm0.0015$ \\
$\Omega_{2}$=$\Omega_{in}$ & 2.8045               & $b$                \\
$X_{1}, X_{2}(bolo)$                  & 0.654, 0.646     & $c$                \\ 
$Y_{1}, Y_{2}(bolo)$                  & 0.119, 0.211      & $c$                \\ 
$x_{1}, x_{2}$($B$)                    & 0.768, 0.837     & $c$                \\ 
$y_{1}, y_{2}$($B$)                    & 0.331, 0.140     & $c$                \\ 
$x_{1}, x_{2}$($V$)                    & 0.660, 0.760     & $c$                \\ 
$y_{1}, y_{2}$($V$)                    & 0.281, 0.234     & $c$                \\ 
$L_{1}/(L_{1}+L_{2})(B)$            & 0.9354              & $\pm0.0005$  \\
$L_{1}/(L_{1}+L_{2})(V)$            & 0.8971              & $\pm0.0006$  \\
$r_{1}$(pole)                              & 0.3989              & $\pm0.0002$  \\
$r_{1}$(point)                             & 0.4732              & $\pm0.0005$  \\
$r_{1}$(side)                              & 0.4200              & $\pm0.0002$  \\
$r_{1}$(back)                             & 0.4406              & $\pm0.0003$  \\                           
$r_{2}$(pole)                              & 0.2938              & $\pm0.0002$  \\
$r_{1}$(point)                             & 0.4216              & $\pm0.0002$  \\
$r_{2}$(side)                              & 0.3065              & $\pm0.0002$  \\
$r_{2}$(back)                             & 0.3391              & $\pm0.0002$  \\
$\sum(O-C)^{2}$     & \multicolumn{2}{c}{0.1797}   \\
\enddata
\tablecomments{ $b$: Assumed; $c$: \citealp{vanh93}.}                                                                                                                    
\end{deluxetable}

\begin{figure}  %% Figure6
\centering
\includegraphics[width=0.48\textwidth, angle=0]{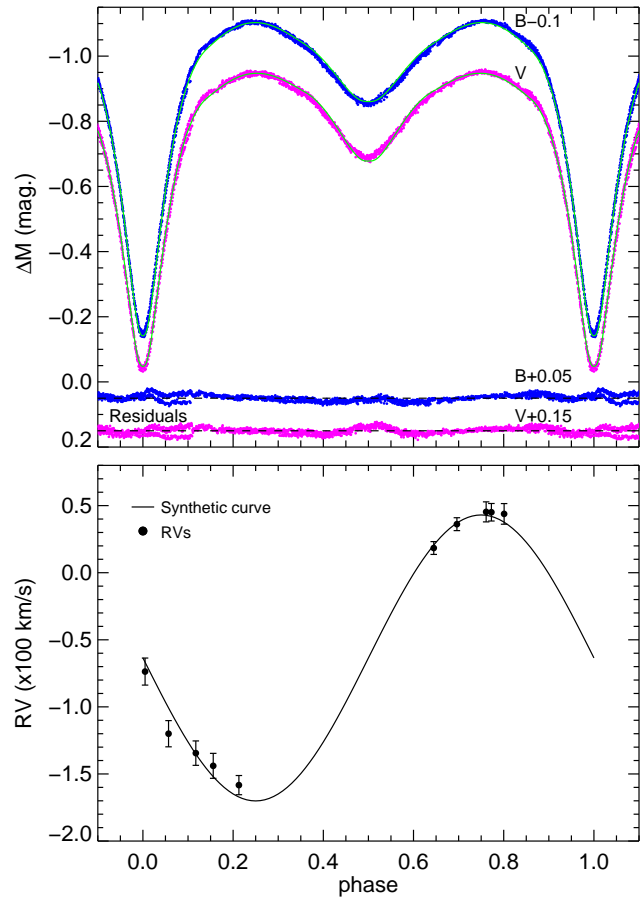}
\caption{The observed $B$ and $V$ band light curves of V Tri and the synthetic curves (green lines) in the up panel.  The observed RVs (filled circles) of the primary component of V Tri and the synthetic (solid line) curve are shown in the down panel.}
\label{fig6}
\end{figure}

The $B$ and $V$ band observed and synthetic light curves are displayed in the up panel of Figure \ref{fig6}. The characteristics of the light curves indicate that this system is a typical Algol-type eclipsing binary system having a semi-detached configuration with the secondary component accurately filling its limiting lobe. Using the equation $f=R/r_{cr}$, where $R$ is the stellar radius and $r_{cr}$ is the critical Roche lobe, we determined that the filling factor of the primary is 93.9\%. It indicates that V Tri is a near-contact semi-detached binary system. The sinusoidal synthetic curve and the observed RVs of the primary component are shown in the down panel of Figure \ref{fig6} and the corresponding sum of squares of the residuals is 0.0622. The systemic radial velocity and the amplitude of RVs are -64 km/s and 107km/s, respectively. Combining the photometric solution with the spectroscopic results, the absolute parameters including the mass, radius, and luminosity for each component of the system were computed from WD code as listed in Table \ref{tab6}. Based on the Kepler's third law $m_1+m_2=(4\pi^2/G)(a^3/P^2)$, the relative radius formula $r=R/a$ and the luminosity $L=4\pi\sigma R^2T^4$, the errors of these parameters are calculated by the error propagation method.

\begin{deluxetable}{lrr}  %%Table6

%Table 6
\tablecaption{Absolute Parameters of V Tri.\label{tab6}}
\tablewidth{700pt}
\tabletypesize{\scriptsize}
\tablehead{ \colhead{Parameter}  &  \colhead{Primary}   &   \colhead{Secondary}}
\startdata
Mass(M$_\odot$)                         &   1.60$\pm$0.07       & 0.74$\pm$0.02             \\
Radius(R$_\odot$)                       &   1.64$\pm$0.02       & 1.23$\pm$0.02             \\
Luminosity(L$_\odot$)                 &  14.14$\pm$0.73       & 1.65$\pm$0.05             \\
\enddata
\end{deluxetable}

%% --Section5
\section{Summary and Discussions} 
\label{sect:5}

We performed time-series photometric observations for the eclipsing binary system V Tri in the $B$ and $V$ bands. Six times of primary and secondary minima were obtained on 2003 December 25, 26, and also from 2007 October 12 to 14. Meanwhile, the light curves were recorded with the CCD camera in a complete phase. We carried out the high-resolution spectroscopic observations with the echelle spectrometer on 2015 November 3 and 5. The spectral type of the system is determined as A3V by comparison with the spectral model in the stellar spectral library by \cite{pick98}. Moreover, ten radial velocities of the primary were determined from the spectra.

We calculated a new linear and quadratic ephemerides for V Tri based on the 81 minimum times that measured by photoelectric and CCD photometry from publications and this paper. The period of this system (0.58520481 days) is very similar to the results of previous research. The orbital period of the system is decreased with the rate of  $-7.80\times10^{-8} d\ yr^{-1}$ over the past few decades. If the decrease of orbital period is attributed to the mass transfer between the two components of the system, the more massive star will transfer mass to the companion. However, a semi-detached configuration with the less massive filling its Roche lobe were yielded by the light curve simulation for the system. It suggests that V Tri used to be an EB-type binary system, the component reached its Roche lobe and rapidly transfers the mass to the primary which was the less massive, the orbital period was decreasing. The system did not exhibit periodical change in X-ray intensity \cite{shaw00}, it indicates that the mass transfer has become inactive. As the mass transfer slows, the accretion disk settle onto the primary that lead to the primary became a relatively normal more massive star. The tidal forces synchronize the rotation of the primary with the orbit and the system is acting as an Algol-type binary from our study. Now V Tri evolve in a slow mass-transfer evolutionary stage on the nuclear timescale. The rate of conservative mass transfer from the component to the primary was estimated to be about $6.12\times10^{-8} M_\odot\ yr^{-1}$.

\citet[Fig.3]{guro06} analysis the $O-C$ residuals that derived from the linear equation based on the collected light minimum times, the light-time effect due to an third body in the binary system was used to explain the result of sinusoidal fit to the residuals. More new photoelectric and CCD minimum times were added to verify  the light-time effect, a sinusoidal-like residuals of all light minima ($Epoch > -20\times10^{3}$, Figure \ref{fig4}) indicate this system is likely to have a third companion. However, this result is not very sure because more visual values form the sine-like structure.

The photometric solution indicates that the temperature of the massive star is 2848 K higher than the less one. It indicated that V Tri could be a A-type near-contact system. The mass ratio and orbital inclination of V Tri were derived to be $q=0.463\pm0.001$ and $84.20\pm0.05$, which are slightly larger than the ratio of $0.458\pm0.005$ and the inclination of $84.07\pm0.19$ given by \cite{guro06}. The absolute parameters for each components of the binary system were computed from the photometrical results as given in Table \ref{tab6}, these parameters are also very different from the previous values. Thanks to the high precision time-series photometrical data and high-resolution spectroscopic data, our credible parameters will help to better understand the character of V Tri at present evolutionary stage.

%%Acknowledgments
\section*{Acknowledgments}

JNF acknowledges the support from the National Natural Science Foundation of China (NSFC) through the grant 11673003 and the National Basic Research Program of China (973 Program 2014CB845700 and 2013CB834900). LFM and RM acknowledge the financial support from the UNAM under grant PAPIIT IN 105115. XBZ and YPL would like to acknowledge partial supported by the NSFC and the NSFC/CAS Joint Fund of Astronomy through grants 11373037 and 11303021. New CCD photometric observations of V Tri were obtained with 85 cm telescopes at the Xinglong Observation Base in China. Based upon spectroscopic observations carried out at the Observatorio Astron\'omico Nacional on the Sierra San Pedro M\'artir (OAN-SPM), Baja California, M\'exico. We thank the daytime and night support staff at the OAN-SPM and Xinglong Observation Base for facilitating and helping obtain our observations. We thank Raul Michel for his help in applying for observation time of the 2.12m telescope at SPM. The authors are grateful to the referee for helpful comments and suggestions on the manuscript.

%%Bibliography

\end{document}